\documentclass[10pt]{article}
\input epsf
\begin{document}

\title{Jeans Instability in the linearized Burnett regime}
\author{A. Sandoval-Villalbazo$^a$ and L.S. Garc\'{\i}a-Col\'{\i}n$^{b,\,c}$ \\
$^a$ Departamento de F\'{\i}sica y Matem\'{a}ticas, Universidad Iberoamericana \\
Lomas de Santa Fe 01210 M\'{e}xico D.F., M\'{e}xico \\
E-Mail: alfredo.sandoval@uia.mx \\
$^b$ Departamento de F\'{\i}sica, Universidad Aut\'{o}noma Metropolitana \\
M\'{e}xico D.F., 09340 M\'{e}xico \\
$^c$ El Colegio Nacional, Centro Hist\'{o}rico 06020 \\
M\'{e}xico D.F., M\'{e}xico \\
E-Mail: lgcs@xanum.uam.mx} \maketitle
\bigskip

\begin{abstract}
Jeans instability is derived for the case of a low density
self-gravitating gas beyond the Navier-Stokes equations. The Jeans
instability criterium is shown to depend on a Burnett coefficient
if the formalism is taken up to fourth order in the wave number.
It is also shown that previously known viscosity corrections to
the Jeans wave-number are enhanced if the full fourth order
formalism is applied to the stability analysis.
\end{abstract}

\bigskip

\section{Introduction}

The physics of low density systems is relevant in the description
of thermodynamic processes which presumably occurred in the
universe some time between its age of $10^{6}$ sec, temperature
$T=1$ KeV and the stage after matter decoupled from radiation when
the matter density reached values of $\sim 10^{-18}
\frac{g}{cm^{3}}$  and $T=1$ eV. It is also known that these types
of regimes prevail in many other astrophysical systems
$\cite{cero}$. Since a hydrodynamical, Navier-Stokes, description
of a low density fluid breaks down for small Knudsen numbers, it
is desirable to go beyond this regime and analyze the fluid with
the tools that statistical physics provide as the particle
collision frequency decreases. One simple tool useful for this
purpose is, on the one hand,  the phenomenological formalism now
known as Linear Irreversible Thermodynamics (LIT), which provides
a good grasp about how non-equilibrium processes evolve in time
\cite{Groot}. On the other hand, kinetic theory provides a solid
mesoscopic foundation for this theory, specially if we want to
deal with hydrodynamics beyond the Navier-Stokes regime \cite{GC}
\cite{Cow}.

The relevance of transport coefficients in cosmological and
astrophysical problems has already been widely recognized
\cite{cero} \cite{Spiegel}  \cite{Weinberg}. Nevertheless, the
Burnett equations \cite{GC} \cite{Cow} have rarely been applied to
structure formation problems in a phenomenological context,
although the Burnett regime itself has been addressed in the
context of transport theory in special relativity \cite{Anderson1}
\cite{Anderson2} \cite{Cercignani}. One possible motivation for
the study of the Burnett regime in cosmological situations is the
well-known relation between wave scattering processes, density
fluctuations and dissipative effects \cite{Berne} \cite{Mountain}
\cite{Yip}. Dynamic structure factors, possibly relevant for
describing CMB distortions, can be derived from the hydrodynamical
Burnett approach, as it has already been done for collisionless
plasmas \cite{yo}. The use of these structure factors has appeared
to be useful in the study of scattering laws to describe the
Sunyaev-Zel'dovich effect and can, in principle, be applied to
study density-density correlation functions to deal with
hydrodynamical instabilities \cite{pa} or temperature-temperature
correlation functions, a problem that remains to be studied via a
thermodynamical approach. For the background microwave radiation
this is done at present using the well-known multipole expansion.
In this work we analyze the solutions to the phenomenological
linearized Burnett equations in the context of Jeans instability
theory, generalizing previous work \cite{Corona} \cite{Mimismo}.
The main reason for doing this is that the introduction of the
Burnett regime allows to take into account all $k^{4}$ order terms
in the dispersion relation. Furthermore, the general tenets behind
this formalism are worth emphasizing in view of its wide range of
applicability \cite{Reese}. To accomplish this task, the paper is
divided as follows: section 2 reviews the basic phenomenological
equations governing the Burnett regime. Section three is dedicated
to the analysis of the dispersion relation obtained form the
linearized formalism. Final remarks about the implications of the
Burnett regime in structure formation are included in the last
section.

\section{Basic formalism}

The starting point for our calculation concerns the method whereby
for a simple fluid which is not in equilibrium we can apply  the
adequate information to transform the well-known conservation
equations for mass, momentum and energy into a complete set. For
this purpose, we assume that the fluid is isothermal and,
therefore, the energy equation may be ignored. This leaves us with
the continuity and momentum equations. The second one has the
form,

\begin{equation}
\label{cero} \rho \frac{du^{i}}{dt}+ \frac{ \partial \tau^{i
j}}{\partial x^{j}}=f^{i}
\end{equation}
where $f^{i}$  is the external force per unit of volume. To relate
the stress tensor $\tau^{i j} $  with the mass density $\rho$ and
the local velocity $u^{i}$, one must appeal to experiment or to
kinetic theory. The main problem here is that Eq. (\ref{cero})
together with the continuity equation involve $4$ variables,
$\rho$ and $u^{i}$. Thus, there are four equations but ten
unknowns assuming that the fluid is isotropic and, therefore,
$\tau^{i j} $ is symmetric. So, we require knowledge of the
constitutive equations relating $\tau^{i j}$ to the set of local
variables. When this procedure is accomplished, keeping all those
terms which are at most quadratic in the gradients of $\rho$ and
$u^{i}$ we refer to it as the Burnett regime. How to accomplish
this using kinetic theory is now briefly described.  We assume
that the fluid is a dilute inert gas whose dynamics is governed by
the Boltzmann equation \cite{GC} \cite{Cow}. For such a model, it
is well known that through the Chapman-Enskog method one can
obtain the solutions to the equation which are consistent with the
macroscopic hydrodynamic equations. This is achieved by expanding
the single particle distribution function in power series of the
Knudsen parameter $\epsilon$, which is a measure of the spatial
gradients present in the gas, responsible for its deviations from
the equilibrium states. When this procedure is carried over, one
obtains, to order zero in $\epsilon$ the Euler regime, and to
first order in $\epsilon$ the Navier-Stokes regime with zero bulk
viscosity, so that
\begin{equation}\label{cerouno}
\tau^{i j}_{NS}=p \delta^{i j} -2\eta(\sigma^{i j}-\frac{1}{3}
\theta \delta^{i j} )
\end{equation}
where $\sigma^{i j}$ is the symmetric traceless part of the
velocity gradient, $\eta$ is the shear viscosity, $p$ is the local
pressure, $\delta^{i j}$ is Kronecker's delta and $\theta
\equiv\frac{\partial u^{i}}{\partial x^{i}}$. To next order in
$\epsilon$ one gets the Burnett regime, in which $\tau^{i j}$
becomes a rather complicated expression \cite{Cow} containing
both, nonlinear as well as linear terms in the velocity gradient.
If we further assume that the gas is slightly deviating form its
local equilibrium state, the \emph{linearized} Burnett
contribution to the stress tensor $\tau^{i j}_{B}$ is then given
by:
\begin{equation}\label{cerotres}
\tau^{i j}_{B}= \varpi _{2} \frac{\eta^{2}}{p_{o}}[\frac{
\partial (\frac{1}{\rho_{o}}f^{i})}{\partial x_{j}} -
\frac{1}{\rho_{o}}\frac{\partial}{\partial x_{j}} (\frac{\partial
p}{\partial x_{i}})-\frac{1}{3} (\frac{
\partial (\frac{1}{\rho_{o}}f^{i})}{\partial x_{j}}-\frac{1}{\rho_{o}}
\nabla^{2} p ) \delta^{i j}]
\end{equation}
In this equation  $\varpi _{2}$ ia a  transport coefficient whose
value is known for rigid spheres and Maxwell molecules as well,
$\rho_{o} $ and $p_{o}$ are the equilibrium values for the density
and pressure, respectively and $p$ the local pressure
$p(x^{i},t)$. $\tau^{i j}$ in Eq. (\ref{cero}) is the total stress
tensor, $\tau^{i j}_{NS}+\tau^{i j}_{B}$, where $\tau^{i j}_{NS}$
is given in Eq. (\ref{cerouno}). When we add these contributions,
take their divergence and use the local equilibrium assumption, $
p=p(\rho,T=const)$, so that $ \frac{\partial p}{\partial
x^{i}}=\frac{\partial p}{\partial \rho} \frac{\partial
\rho}{\partial x^{i}}=C_{s}^{2} \frac{\partial \rho}{\partial
x^{i}}$ one finally gets that the linearized Burnett approximation
to Eq.(\ref{cero}) is given by:

\begin{eqnarray}
\rho_{o} \frac{\partial u^{i} }{\partial t}+
C_{s}^{2}\frac{\partial (\delta \rho)}{\partial x_{i}} -2 \eta
(\nabla ^{2}u^{i}-\frac{1}{3} \nabla_{i} (\nabla \cdot \vec{u}))
\nonumber \\
+\frac{\varpi _{2} \eta^{2}}{p_{o}}[\nabla^{2}
(\frac{f^{i}}{\rho_{o}})-\frac{1}{3} \nabla_{i} \nabla \cdot
(\frac{\vec{f}}{\rho_{o}})-\frac{2}{3 \rho_{o} }C_{s}^{2}
\nabla^{2} \nabla_{i} (\delta \rho)]=f^{i} \label{one}
\end{eqnarray}
where $C_{s}^{2}$ is the speed of sound.

The linearization procedure, which is identical to the one
followed in a previous publication \cite{pa}, has clearly invoked
that local state variables $\rho$ and $u^{i}$ deviate form its
equilibrium values through linear fluctuations, namely
\begin{equation}\label{f1}
\rho=\rho_{o}+\delta \rho
\end{equation}
and
\begin{equation}\label{f2}
u^{i}=u^{i}_{o}+\delta u^{i}=\delta u^{i}
\end{equation}
The continuity equation now reads as
\begin{equation}
\frac{\partial (\delta \rho )}{\partial t}+ \delta \theta =0
\label{oneb}
\end{equation}
where $\delta \theta=\frac{\partial u^{i}}{\partial x^{i}} $. Eqs.
(\ref{one})-(\ref{oneb}) constitute the set of linearized coupled
equations for $\delta \rho$ and $u^{i}$ which can  be solved for a
given conservative force $f^{i}$. In our case of interest, the
external force density  is taken to be the gravitational force
and, moreover, we assume that the fluctuating part of the
gravitational field $\delta \varphi $ is governed by Poisson's
equation
 \begin{equation}\label{twob}
 \nabla^{2} \delta \varphi=4 \pi G \delta \rho
\end{equation}
where $G=6.67 \times 10^{-11} \frac{m^{3}}{kg s^{2}}$ is the
gravitational constant.

To eliminate $u^{i}$ from this set we simply take the divergence
of Eq. (\ref{one}) and make use of Eq. (\ref{twob}) plus the fact
that $\frac{\partial}{\partial t} \nabla \cdot =\nabla \cdot
\frac{\partial}{\partial t} $ so that after a straightforward
arrangement of terms we find that
\bigskip
\begin{eqnarray}
 \frac{\partial ^{2}\left( \delta \rho \right) }{\partial
t^{2}} -C_{s}^{2}\nabla ^{2}\left( \delta \rho \right) - \frac{8
\pi }{3 p_{o}} \varpi_{2} \eta^{2} G \nabla^{2} \left(
\delta \rho \right) \nonumber \\
-\frac{4}{3}\frac{\eta}{\rho_{o}}\nabla^{2}\left( \frac{\partial
\left( \delta \rho \right)}{\partial t} \right) +\frac{2}{3}
\varpi_{2} (\frac{\eta ^{2}}{p_{o} \rho_{o} }C_{s}^{2})
\nabla^{2}(\nabla^{2} \delta \rho)+4\pi G\rho_{o}\,\delta \rho =0
\label{doce}
\end{eqnarray}
Eq. (\ref{doce}) is the main result in this calculation. It is a
single equation describing the density fluctuations of an
isothermal fluid when subject to the action of a fluctuating
gravitational field taking into account all linearized
contributions up to second order in the Knudsen parameter. Its
implications on the Jeans number will be studied in the following
section. Before we do so, however, it is interesting to point out
the main features that in Eq. (\ref{doce}) originate from the
Burnett approximation, those proportional to $\eta^{2}$. The one
of fourth order in the spatial derivatives, the next to the last
term in the left hand side, corresponds to what was to be
expected. But the other one namely, the third term in the left
hand side shows a coupling between the gravitational field and the
dynamics of the density fluctuations. This coupling is totally
absent both at the Euler and the Navier-Stokes level. Its
implications will be discussed later.

\section{Analysis of the dispersion relation}

In order to solve  Eq. (\ref{doce}) we propose that the density
fluctuations are represented by plane waves, so that
\begin{equation}
\delta \rho \left( \vec{r},t\right) =Ae^{-i(wt-\vec{k}\cdot
\vec{r})} \label{trece}
\end{equation}
where the amplitude $A$ needs not to be specified and the rest of
the symbols have their standard meaning. Substitution of Eq.
(\ref{trece}) into Eq. (\ref{doce}) yields:

\begin{equation}
\omega^{2}+k^{2}(-C_{s}^{2}+\frac{8 \pi}{3 p_{o}} \varpi_{2}
\eta^{2} G \rho_{o})+\frac{4}{3}\frac{\eta}{\rho_{o}} i \omega
k^{2}-\frac{2}{3}\varpi _{2} \frac{\eta^{2} C_{s}^{2}}{p_{o}
\rho_{o}}k^{4}+4 \pi G \rho_{o}=0 \label{catorce}
\end{equation}
Eq.(\ref{catorce}) is the sought dispersion relation from which we
must examine the conditions leading to values of $\omega$,
necessarily complex, giving rise to runaway terms in Eq.
(\ref{trece}). In fact, it may be rewritten as

\begin{equation}
\omega^{2}+2 i a(k) \omega+f(k)=0 \label{quince}
\end{equation}
where
\begin{equation}
a(k)=\frac{2}{3} \frac{\eta}{\rho_{o}}k^{2} \label{dieciseis}
\end{equation}
and
\begin{equation}
f(k)=(\frac{C_{s}^{2}}{\gamma}-\frac{8 \pi}{3 \rho_{o}}G
\varpi_{2} \eta^{2}) -\frac{2}{3} \frac{C_{s}^{2}}{\gamma
\rho_{o}} \frac{\varpi _{2} \eta^{2}}{p_{o}}k^{4}+4\pi G \rho_{o}
\label{diecisiete}
\end{equation}

When examining the roots of this quadratic equation we see that
the runaway solutions require that $\omega$ be complex and the
threshold values for $k$ to meet this requirement is that
$a(k)^{2} +f(k) \geq 0$. This condition reduces to a quadratic
equation for the minimum value of $k^{2}$ namely,
\begin{equation}
r^{2} k^{4}+s k^{2}+4 \pi G \rho_{o}=0 \label{diecisieteb}
\end{equation}
where
\begin{equation}\label{x1}
r^{2}=\frac{4}{9} \frac{\eta^{2}}{\rho_{o}^{2}}-\frac{2}{3}
\varpi_{2} \eta^{2} G \rho_{o}
\end{equation}
and
\begin{equation}\label{x2}
s=-C_{s}^{2}-\frac{8 \pi}{3 p_{o}} \varpi_{2} \eta^{2}
\frac{C_{s}^{2}}{\gamma \rho_{o}}
\end{equation}

Calculating the roots of Eq. (\ref{diecisieteb}), taking the root
with the positive sign, $k^{2}>0$, and expanding the square root
we find, after a few trivial algebraic steps, that

\begin{equation}
\frac{k^{2}}{k_{J}^{2}}=1+\frac{\eta^{2}}{\rho_{o}}k_{J}^{2}(\frac{2
\varpi_2}{p_{o}}-\frac{4}{9}\frac{\gamma}{Cs^{2} \rho_{o}})
\label{veinte}
\end{equation}
where the ordinary Jeans wave number is given by
$k_{J}^{2}=\frac{4\pi G\rho _{o}}{C_{s}^{2}}$. Eq.(\ref{veinte})
is the sought result. In its derivation we have consistently
neglected all terms of order $\eta^{4}$ and higher.

To what extent this result has an influence on the ordinary Jeans
number will be discussed below. Two minor points should be
noticed. Firstly, if $\varpi_{2}=0$ the corrective term, which is
of order of $k^{4}$ is the one obtained within the Navier-Stokes
approximation when ignoring the bulk viscosity  \cite{Mimismo}.
Secondly, as we have pointed out initially, the Burnett linearized
terms have to be included to have a consistent correction to order
$k^{4}$.

\section{Discussion}

As Eq. (\ref{veinte}) clearly indicates, the inclusion of the
Burnett linearized term provides a contribution to the correction
of $k_{J}^{2}$ which is opposite in sign to the one arising form
the Navier-Stokes equations. This may have a significant effect on
the Jeans mass, which is inversely proportional to $k_{J}^{3}$
\cite{Weinberg} \cite{yo} \cite{Corona}.

One may now ask about the conditions in which dissipative effects
are indeed relevant in structure formation problems. In order to
address this question we start with  typical monatomic gases such
as H, He, Ar, etc which we know, are reasonably described by a
hard-sphere model whose shear viscosity extracted from kinetic
theory is given by \cite{Cow}:

\begin{equation}
\eta = \frac{5}{16 \sigma^{2}} \sqrt{\frac{k_{B} m
T}{\pi}}\label{veintiseis}
\end{equation}
Here $ \sigma $ is the atomic diameter, $ m $ the mass of the
particles, $ k_{B}$ is Boltzmann's constant and $ T $ is the
temperature of the system. The most relevant contribution will
arise from the term within brackets. To estimate its value we
recall that for hard spheres $\varpi_{2}\simeq 2$ and the equation
of state is $p_{o}=\frac{\rho_{o} R T}{M} $, where $R$ is the
universal gas constant and $M$ the molecular weight of the gas.
Then, the correction is determined by the expression

\begin{equation}
\hat{C} \equiv \frac{k^{2}-k_{J}^{2}}{k_{J}^{2}}=\frac{4 \eta^{2}
k_{J}^{2}}{\rho_{o}}(\frac{2\varpi_{2}}{p_{o}}-\frac{\gamma}{9
C_{s}^{2} \rho_{o}}) \label{veintisiete}
\end{equation}

Using the equation of state, the value of $\varpi_{2}$ and the
fact that for a monatomic gas $C_{s}^{2}=\frac{5}{3} \frac{k_{B} T
}{M m_{H}}$, then for hydrogen $M=2$ whence Eq.(\ref{veintisiete})
reduces to
\begin{equation}
\hat{C} = \frac{8 \eta^{2} k_{J}^{2}}{R \rho_{o} T}(1-\frac{
\gamma}{15}) \label{veintisieteb}
\end{equation}

Since $\gamma \sim 1.7 $ for monatomic gases, this equation
finally reads as
\begin{equation}
\hat{C} \simeq \frac{5}{8} \frac{G M}{\sigma^{4}}
\frac{m_{H}^{2}}{R \rho_{o} T}\cong \frac{5}{\rho_{o} T}\times
10^{-19}\label{veintisietec}
\end{equation}

Eq. (\ref{veintisietec}) has to be taken as a result which
requires some care before conclusions may be extracted from it. As
we said in the introduction, the Burnett regime is an adequate
formalism to deal with fluids in the so called \emph{transition
regime}. This implies the regime prevailing between the rarefied
regime and the continuous one. In the latter, the well-known
Navier-Stokes equations of hydrodynamics provide an accurate
description of the fluid. It appears that this transition regime
could have prevailed in the Universe for temperatures $T \sim
10^{4}$ (K) and densities $10^{-15}$ $\frac{kg}{m^{3}}$, but these
figures are merely qualitative. If such conditions existed, then
Eq. (\ref{veintisietec}) simply asserts that structure formation
would have been enhanced due to the coupling between hydrodynamic
and gravitational modes. Notice that this coupling is the dominant
term of Eq. (\ref{veintisieteb}). Clearly the importance of the
correction critically depends on $\rho_{o}$ and $T$. If $\rho_{o}
T $ is larger than $ 10^{-19}$ as it probably occurred in earlier
stages, the correction $\hat{C}$ will be negligible so that, in
order to reach a more definite statement, more reliable data is
required. As a final comment, it should be mentioned here that in
this calculation we have ignored taking into account the expansion
in a comoving reference frame. Whether or not this improves the
results is a subject of further study.

This work has been partially supported by CONACyT (Mexico),
project 41081-F.

\end{document}